\documentclass[conference]{IEEEtran}
\usepackage{graphics}
\usepackage[dvips,pdftex]{graphicx}
\usepackage{amsmath,epstopdf,subfig,caption}
\usepackage{flushend}
\usepackage{multirow}

\def \sys {\textit{IncVoronoi}}
\newcommand{\rulesep}{\vrule
}

\usepackage{cite}

\ifCLASSINFOpdf
\else
\fi

\newcommand\blfootnote[1]{%
  \begingroup
  \renewcommand\thefootnote{}\footnote{#1}%
  \addtocounter{footnote}{-1}%
  \endgroup
}
\begin{document}
\title{A Robust Zero-Calibration RF-based Localization System for Realistic Environments}

\author{\IEEEauthorblockN{Rizanne Elbakly}
\IEEEauthorblockA{Wireless Research Center\\
E-JUST, Egypt\\
Email: rizanne.elbakly@ejust.edu.eg}
\and
\IEEEauthorblockN{Moustafa Youssef}
\IEEEauthorblockA{Wireless Research Center\\
E-JUST, Egypt\\
Email: moustafa.youssef@ejust.edu.eg}
}

\maketitle

\begin{abstract}
Due to the noisy indoor radio propagation channel, Radio Frequency (RF)-based location determination systems usually require a tedious calibration phase to construct an RF fingerprint of the area of interest. This fingerprint varies with the used mobile device, changes of the transmit power of smart access points (APs), and dynamic changes in the environment; requiring re-calibration of the area of interest; which reduces the technology ease of use. 

In this paper, we present \sys{}: a novel system that can provide zero-calibration accurate RF-based indoor localization that works in realistic environments. The basic idea is that the relative relation between the received signal strength from two APs at a certain location reflects the relative distance from this location to the respective APs. Building on this, \sys{} incrementally reduces the user ambiguity region based on refining the Voronoi tessellation of the area of interest. 
\sys{} also includes a number of modules to efficiently run in realtime as well as to handle practical deployment issues including the noisy wireless environment, obstacles in the environment, heterogeneous devices hardware, and smart APs.

We have deployed \sys{} on different Android phones using the iBeacons technology in a university campus.
Evaluation of \sys{} with a side-by-side comparison with traditional fingerprinting techniques shows that it can achieve a consistent median accuracy of 2.8m under different scenarios with a low beacon density of one beacon every 44m$^2$. Compared to fingerprinting techniques, whose accuracy degrades by at least 156\%, this accuracy comes with no training overhead and is robust to the different user devices, different transmit powers, and over temporal changes in the environment. This highlights the promise of \sys{} as a next generation indoor localization system.
\end{abstract}

\IEEEpeerreviewmaketitle

\section{Introduction}
\blfootnote{Moustafa Youssef is currently on sabbatical from Alexandria University, Egypt.}
The widespread use of wireless networks combined with ubiquitous mobile devices led to the proliferation of RF-based localization techniques \cite{ips_vision,youssef2005multivariate,el2010propagation,abdel2013monophy,sabek2012multi,youssef2006location,ibrahim2011hidden,kosba2009analysis,saeed2014ichnaea}. Such techniques, e.g. WiFi-based localization and the more recent iBeacons-based localization that leverages the Bluetooth Low Energy (BLE) technology, provide a solution for ubiquitous indoor tracking. Typically, RF-based indoor localization systems require a tedious calibration phase of the area of interest, where an RF ``fingerprint'' is constructed to capture the noisy and complex indoor propagation channel. This fingerprint requires continuous updates to capture temporal changes in the environment and dynamic changes of the transmit power of the smart access points (APs), e.g. to handle interference 
on the AP. Moreover, the heterogeneity of user devices introduces another challenge of capturing the hardware differences between these different devices in the fingerprint.

To address these challenges, a number of RF localization techniques have been proposed that try to reduce the calibration overhead through crowd-sourcing of the fingerprint construction that require explicit \cite{organic} or implicit \cite{abdelnassersemanticslam,wang2012no,rai2012zee} user feedback, use propagation models to automate the fingerprint construction process \cite{aroma,RADAR00,ARIADNE}, or combine RF localization with other sensors \cite{abdelnassersemanticslam,wang2012no}. These systems, however, trade the accuracy with overhead and usually incur higher energy consumption and/or cost. On the other hand, to handle the devices heterogeneity, a number of approaches that map the fingerprint constructed by one device to another have been introduced \cite{het5,het2,het3,het4}. Nevertheless, the range of available user devices in the market, which keeps growing each day, makes the mapping not accurate in all cases and requires some training/learning process for each new device. Finally, to address the dynamic power changes of smart APs and other temporal changes, special sniffers have been used \cite{lease} to monitor the area of interest, increasing the deployment cost of the localization system. These challenges highlight the need for a system that is calibration-free, accurate, robust to heterogeneity in user devices, and adapts to dynamic changes in the environment and APs transmit power.

\begin{figure}[!t]
\centering
\includegraphics[width=0.77\columnwidth]{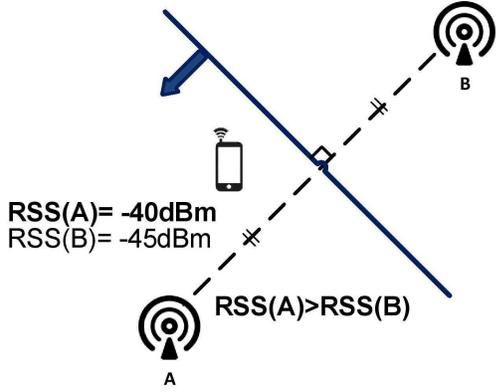}
\caption{\sys{} basic idea (in an \textbf{\emph{ideal}} environment): for any pair of APs, if the RSS from AP $A$ at the device is greater than the RSS from $B$, then the device must be closer to $A$ than $B$. This maps to placing the device in the half plane defined by the bisector line between $A$ and $B$ and containing AP $A$.}
\label{fig:half_plan}
\end{figure}

In this paper, we present the \sys{} system that uses a novel approach to handle the RF indoor localization problem. The basic idea is that, in an ``\textbf{ideal}'' environment, the relative received signal strength (RSS) from the different RF APs\footnote{We use the word AP to refer to any RF transmitter, e.g. a WiFi AP or a
BLE beacon.} at a particular user device can be mapped to the relative distance between the different APs and that device. In other words, the higher the RSS received from an AP, the closer the device should be to this AP. \sys{} leverages this observation by constructing the Voronoi diagram of the area of interest relative to the different AP locations. The initial user ambiguity region is estimated as the Voronoi cell of the strongest heard AP. This region is successively refined by noting that the relative RSS relation between each pair of APs splits the 2D plane of the area of interest into two half-planes (Figure~\ref{fig:half_plan}), further reducing the user ambiguity region. The final user location is estimated as the center of mass of the final user ambiguity  region after applying all RSS constraints. Note that depending on the relative RSS relation, rather than the absolute RSS level, allows \sys{} to handle the user device heterogeneity and reduces the effect of dynamic changes in the environment. \sys{} also incorporates a number of modules to handle \textbf{\emph{practical deployment scenarios that deviate from the ideal scenario}} including handling different and dynamic APs transmit powers, and obstacles in the environment. 
Furthermore, to efficiently calculate the RSS constraints in realtime, \sys{} introduces a gridding approach, where a virtual grid is super-imposed over the area of interest and is used in realtime using a simple comparison operation to significantly reduce the technique's running time and processing power requirements. 

We deployed \sys{} in a university campus using the iBeacons BLE technology over a period of five months with a side-by-side comparison with traditional RF fingerprinting techniques. The results show that \sys{} can achieve a consistent median accuracy of 2.8m under different scenarios. Compared to fingerprinting techniques, whose accuracy degrades by at least 156\%, this accuracy comes with no training overhead and is robust to the different user devices, different iBeacons transmit powers, and over temporal changes in the environment.

The rest of the paper is organized as follows: Section~\ref{sec:basic} presents the \sys{} basic idea in an ideal environment. Section~\ref{sec:system} gives the details of the system and how it handles different practical considerations. We evaluate the system performance in Section~\ref{sec:eval} and compare it to the state-of-the-art. Finally, sections \ref{sec:related} and \ref{sec:conclude} discuss related work and give concluding remarks, respectively.

\begin{figure*}[!t]
    \subfloat[Initial step: User initial ambiguity region is the Voronoi cell corresponding to the AP with the highest RSS.\label{fig:ex_a}]{%
      \includegraphics[width=0.24\linewidth]{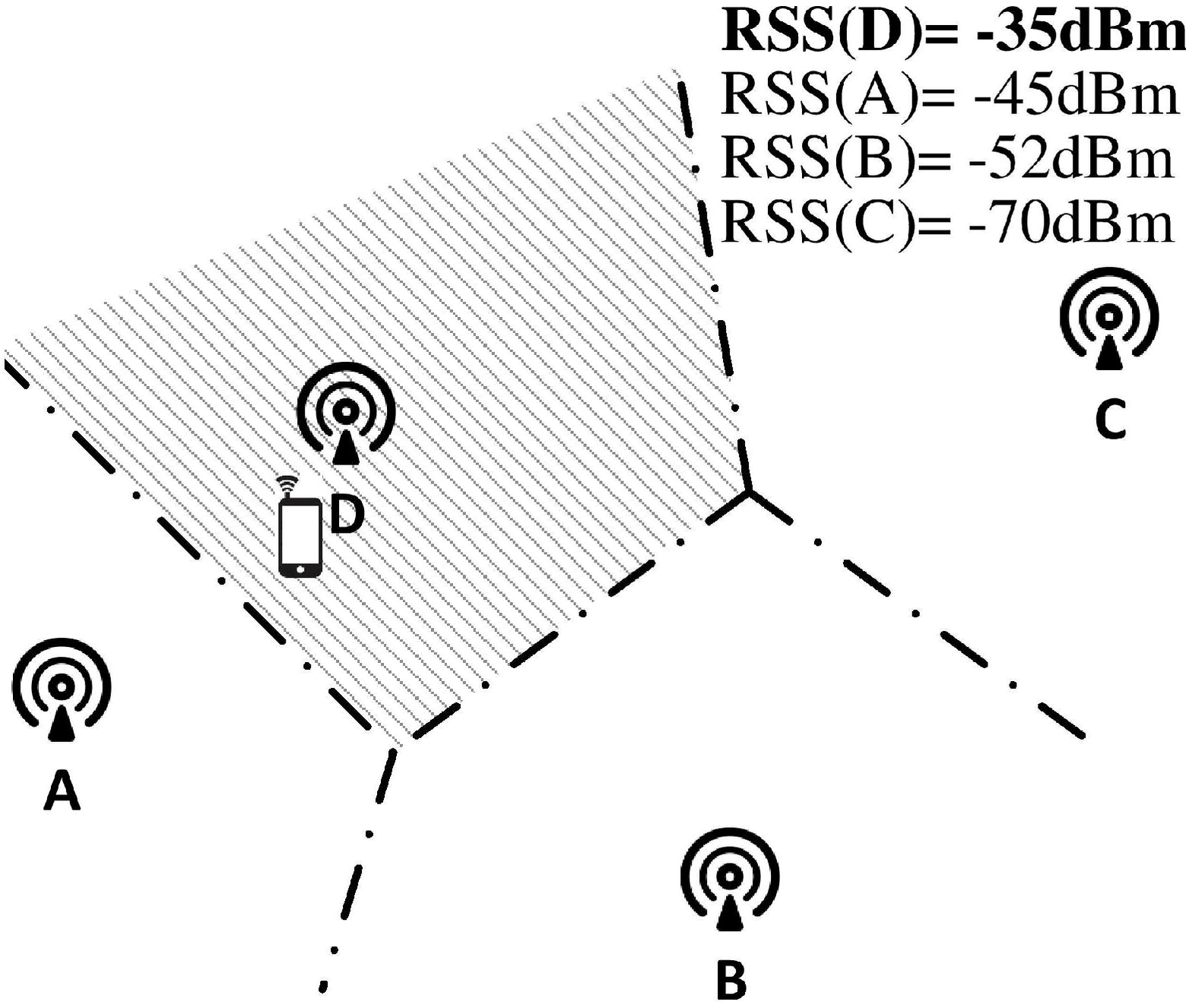}}\hspace{0.08mm}\rulesep
    \subfloat[Applying the first constraint ($B-C$): Since RSS($B$)$>$RSS($C$), the user location falls in the half plane defined by the bisector line between $B$ and $C$ and containing $B$.\label{fig:ex_b}]{%
     \includegraphics[width=0.24\linewidth]{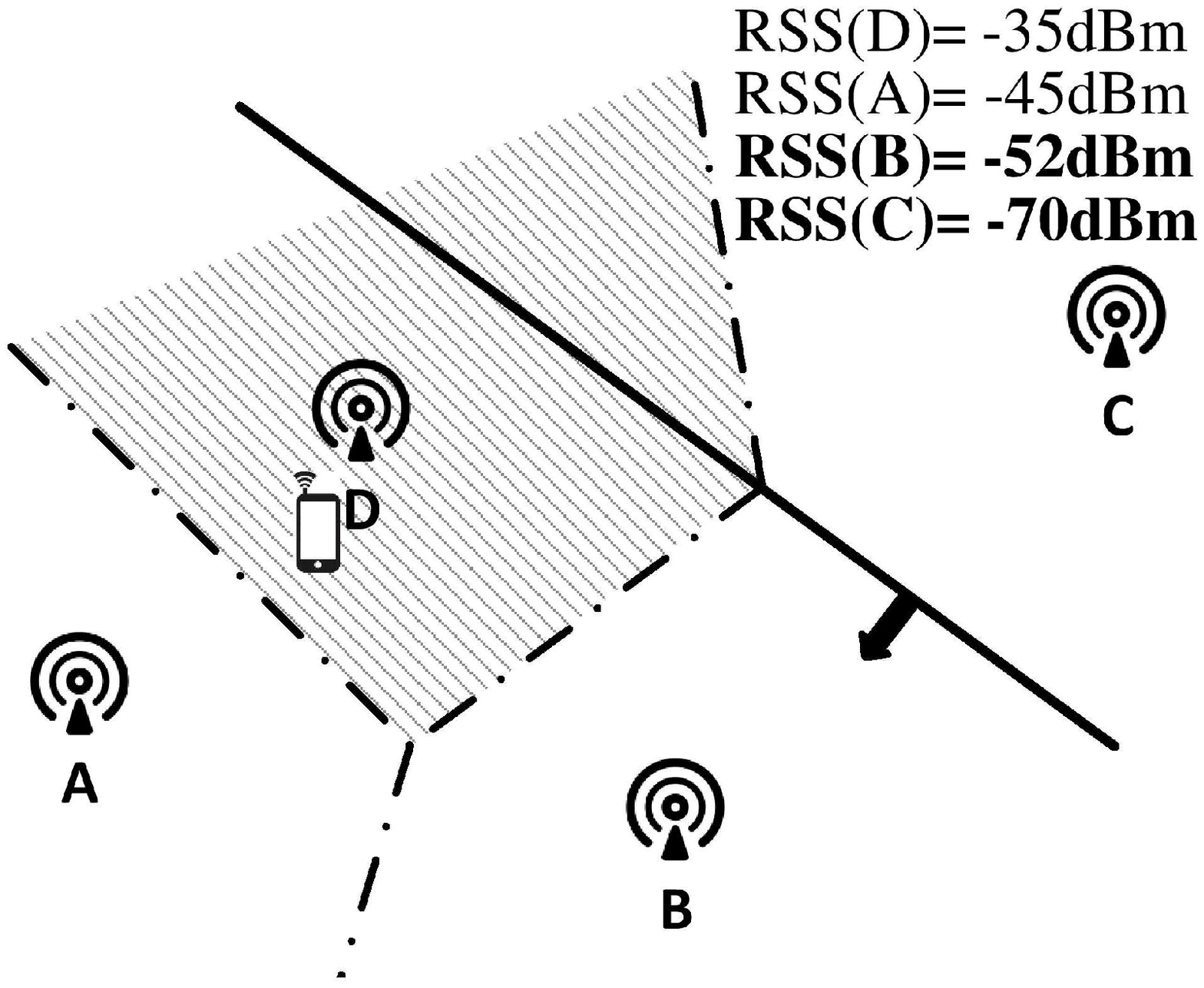}}\hspace{0.08mm}\rulesep
    \subfloat[Applying second constraint ($A-B$): Similarly, the user falls in the half plane defined by the bisector line between $A$ and $B$ and containing $A$. \label{fig:ex_c}]{%
      \includegraphics[width=0.235\linewidth]{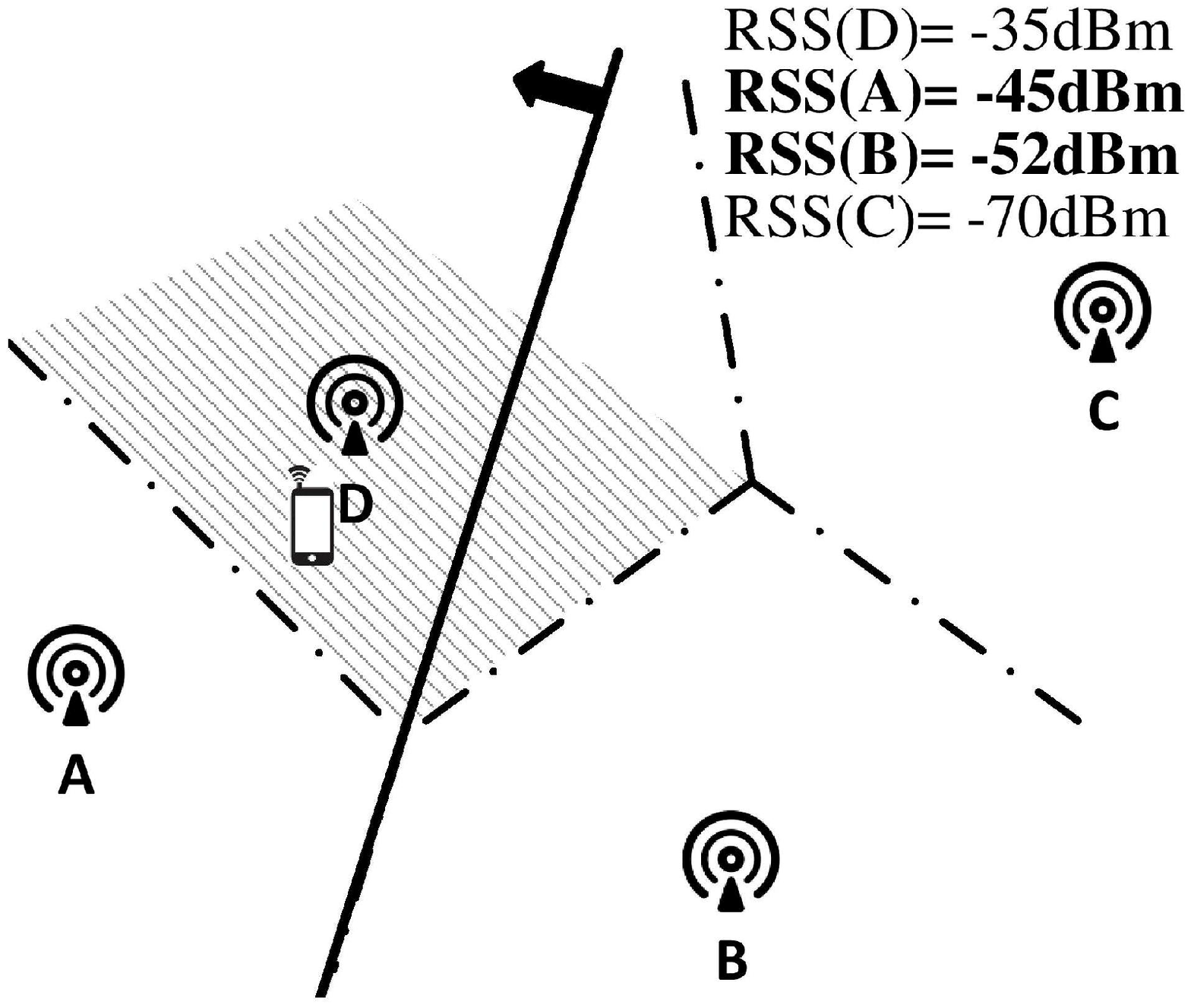}}\hspace{0.08mm}\rulesep
    \subfloat[Final location estimation: The user location is estimated as the center of mass of the last ambiguity region. \label{fig:ex_d}]{%
     \includegraphics[width=0.235\linewidth]{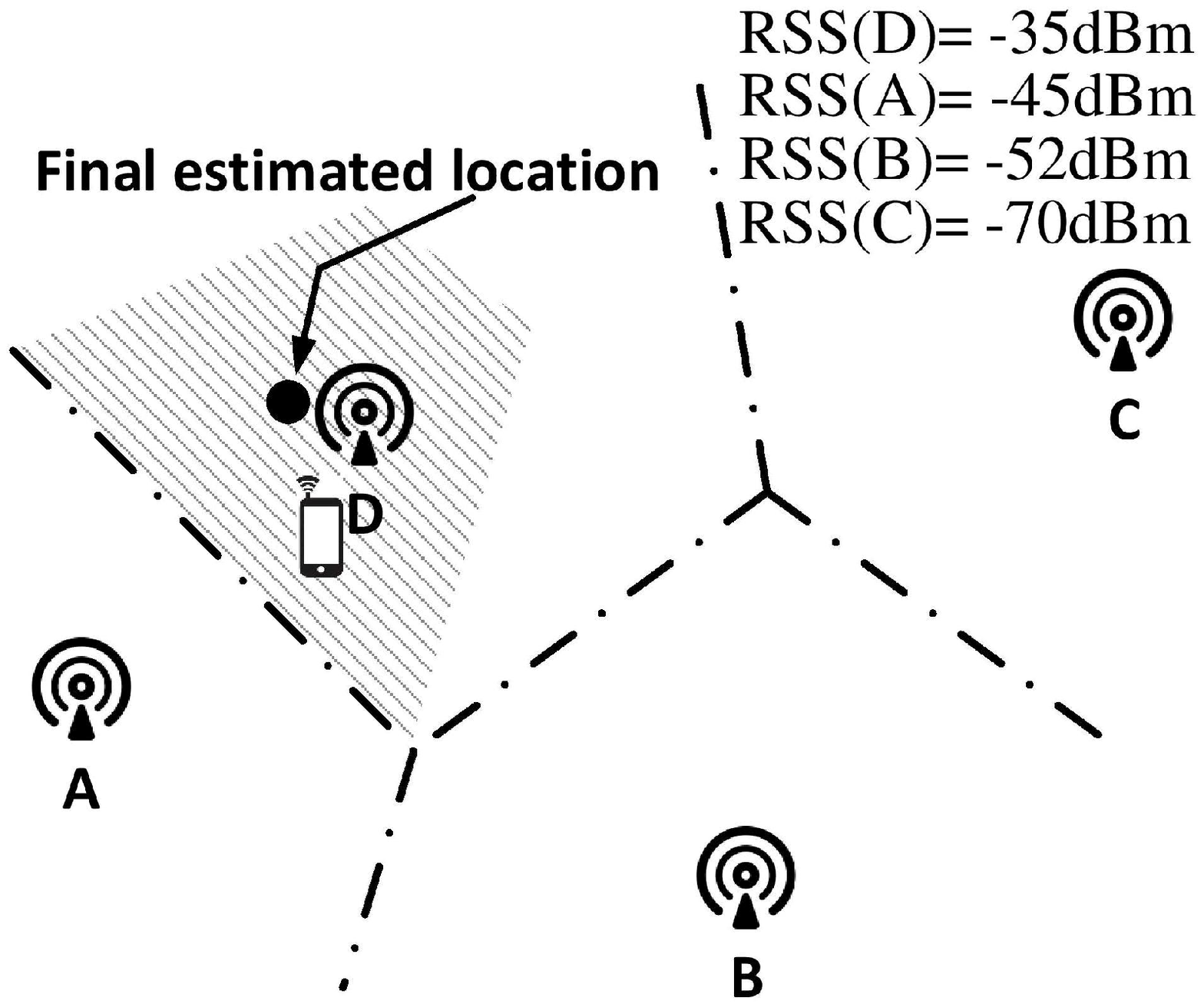}}\hspace{0.08mm}
    \caption{Example of the \sys{} basic approach with four APs. Note that the order of applying the constraints does not matter and that some constraints are redundant given other constraints (e.g. constraint ($A-C$) does not reduce the user ambiguity region after applying the other constraints).}
    \label{fig:example}
  \end{figure*}
	\section{IncVoronoi Basic Idea}
\label{sec:basic}
In this section, we introduce the basic idea \sys{} is built on. We follow this by discussing the different practical challenges that need to be addressed, paving the way for the full system architecture described in the next section.

For ease of explanation, we start by a simple deployment scenario in an open area of interest, i.e. free of obstacles, where all APs 
 are set to transmit with the same power. We relax these assumptions later in this section.
\subsection{Incremental Voronoi Tessellation}
\label{sec:basic_vor}
Figure~\ref{fig:example} shows an example of \sys{} in action for four APs. The basic idea \sys{} builds on is that the relation between the RSS and distance from the transmitter is monotonic. That is, the further away the receiver is from the transmitter, the lower the signal strength and vice versa. Given this, a receiver at a particular location hearing $m$ APs has to be closer to the \textbf{strongest AP} than to the other APs. This is equivalent to placing the receiver at the Voronoi cell of the strongest AP (Figure~\ref{fig:ex_a}) of the Voronoi Tessellation constructed with the APs as seeds \cite{voronoi}.

To further reduce the ambiguity of the user location, \sys{} then applies spatial constraints on the user location based on the pairwise RSS relation of every pair of APs. 
Specifically, given the RSS from two APs RSS$(A)$ and RSS$(B)$, if RSS$(A)$ $>$ RSS$(B)$, then the receiver has to be closer to AP$(A)$ than AP$(B)$ and vice versa (Figure~\ref{fig:half_plan}). In 2D, this maps to the half plane where AP$(A)$ is located. Given that $m$ APs are heard at the current receiver location, we can have up to ${m \choose 2}$ pairwise constraints. Note that some of these constraints define the boundary of the initial Voronoi cell the receiver is located at and some of them are redundant given other constraints.  Incrementally applying all these pairwise constraints further reduces the ambiguity region of the receiver (Figure~\ref{fig:ex_b}-c). The final location of the receiver is estimated as the center of mass of the final ambiguity region (Figure~\ref{fig:ex_d}).
\begin{figure}[!t]
\centering
\includegraphics[width=0.8\columnwidth]{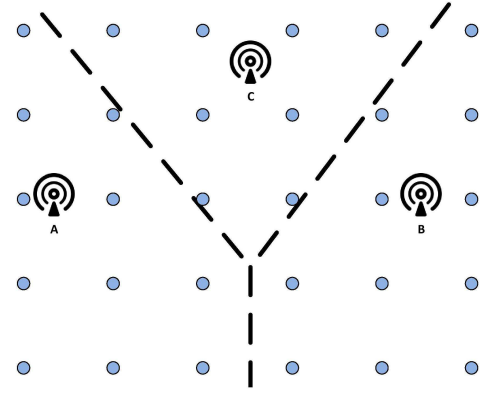}
\caption{\sys{} virtual gridding approach.
} 
\label{fig:vgrid_ill}
\end{figure}
\subsection{Virtual Gridding Approach}
\label{sec:vgrid}
Calculating the constraints in realtime may be computationally intensive to run on battery constrained devices. 
 To address this issue, we propose to use a gridding approach, where a virtual grid is super-imposed on the area of interest (Figure~\ref{fig:vgrid_ill}). For each grid point, the \textbf{\emph{expected}} constraint value for a pair of APs is calculated \textbf{offline} based on the \textbf{\emph{distance}} between the grid point and the different APs. While the system is running in realtime, grid points are scanned and the \emph{\textbf{actual}} constraint value based on the \textbf{\emph{RSS}} by the user device from the two APs is compared to the expected constraint value stored for each grid point. The estimated receiver location is taken as the center of mass of the grid points that have the largest number of matching expected and actual constraints.

Note that the density of the grid points can be chosen to tradeoff the computational complexity and accuracy as we quantify in Section~\ref{sec:eval}. In addition, due to the noisy wireless channel that leads to fluctuations of the RSS values, in some cases some constraints may be conflicting, leading to a null feasible receiver region in the case of the original Incremental Voronoi Tessellation algorithm discussed in the previous section. The Virtual Gridding algorithm described in this section solves this problem by estimating the receiver location based on the  grid points that match the \textbf{\emph{largest number of actual and expected constraints}}.
\subsection{Discussion and Other Practical Considerations}
The presented Incremental Voronoi Tessellation allows one to obtain a calibration-free solution for indoor localization. In addition, it can handle heterogeneous user devices since it only depends on the relative RSS values between pairs of APs, rather than the absolute signal strength.
However, a number of practical considerations still need to be addressed in order to achieve the system goals in practical deployments. These are discussed in this section. The next section provides the \sys{} full system architecture that can handle the following issues:

\subsubsection{Noisy wireless channel}
A receiver stationed at a fixed location will suffer from fluctuation in the RSS due to the noisy wireless channel. These fluctuations may reverse the relative RSS relation between a pair of APs, especially when the signal strengths are close to each other.
\begin{figure}[!t]
    \includegraphics[width=0.77\columnwidth]{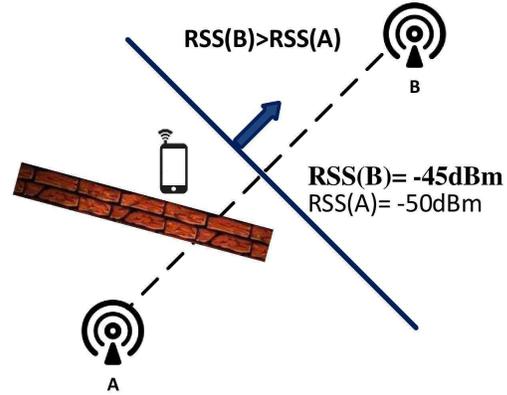}
\caption{A wall between the transmitter $A$ and the receiver leads to reducing the RSS from $A$ (compared to the no wall case in Figure~\ref{fig:half_plan}), reversing the constraint and leading to estimating the device location on the wrong side of the half-plane.}
\label{fig:waf_effect}
\end{figure}
\subsubsection{Different APs transmit powers} 

Modern APs usually contain a feature for automatic power control, where the AP can dynamically change its transmit power to cope with dynamic load or interference. Similarly, different APs can be set to have different transmit powers to provide different coverage goals. This may lead to violating the assumption that the receiver will be closer to the AP with the highest RSS, if the two APs are transmitting with different powers.

\subsubsection{Obstacles}
Similarly, different obstacles in the environment between the transmitter and receiver, especially walls and floors, may significantly reduce the RSS, leading to reversing the constraint for a specific pair of APs (Figure~\ref{fig:waf_effect}). This needs to be handled in a real deployment.

\section{IncVoronoi Full System: Practical Considerations}
\label{sec:system}
In this section, we provide the \sys{} system architecture followed by the details of its modules that handle the different practical considerations and provide an accurate, zero-calibration indoor localization system. 
\subsection{System Model}
Without loss of generality, we assume a 2D area of interest where $n$ APs have been installed. A user carrying a device at an unknown location $l$ scans for the nearby APs. The scanned APs information can be represented as a vector $s=[s_1, ..., s_m]$, where $m\le n$ is the number of heard APs and each $s_i$ is an ordered pair of $(\mathrm{AP}_i, \mathrm{RSS}_i)$ representing the ID and signal strength of the $i^{\textrm{th}}$ heard AP. The entries in the vector $s$ are \textbf{\emph{sorted in a descending order}} according to the RSS.

To reduce the computational overhead, \sys{} uses a virtual gridding approach. Let $g_i$ represent the $i^{\textrm{th}}$ 2D grid point. We also denote the Voronoi cell associated with the $i^{\textrm{th}}$ AP as \emph{Vor}$_i$.

\begin{figure}[!t]
\centering
\includegraphics[width=3.4in]{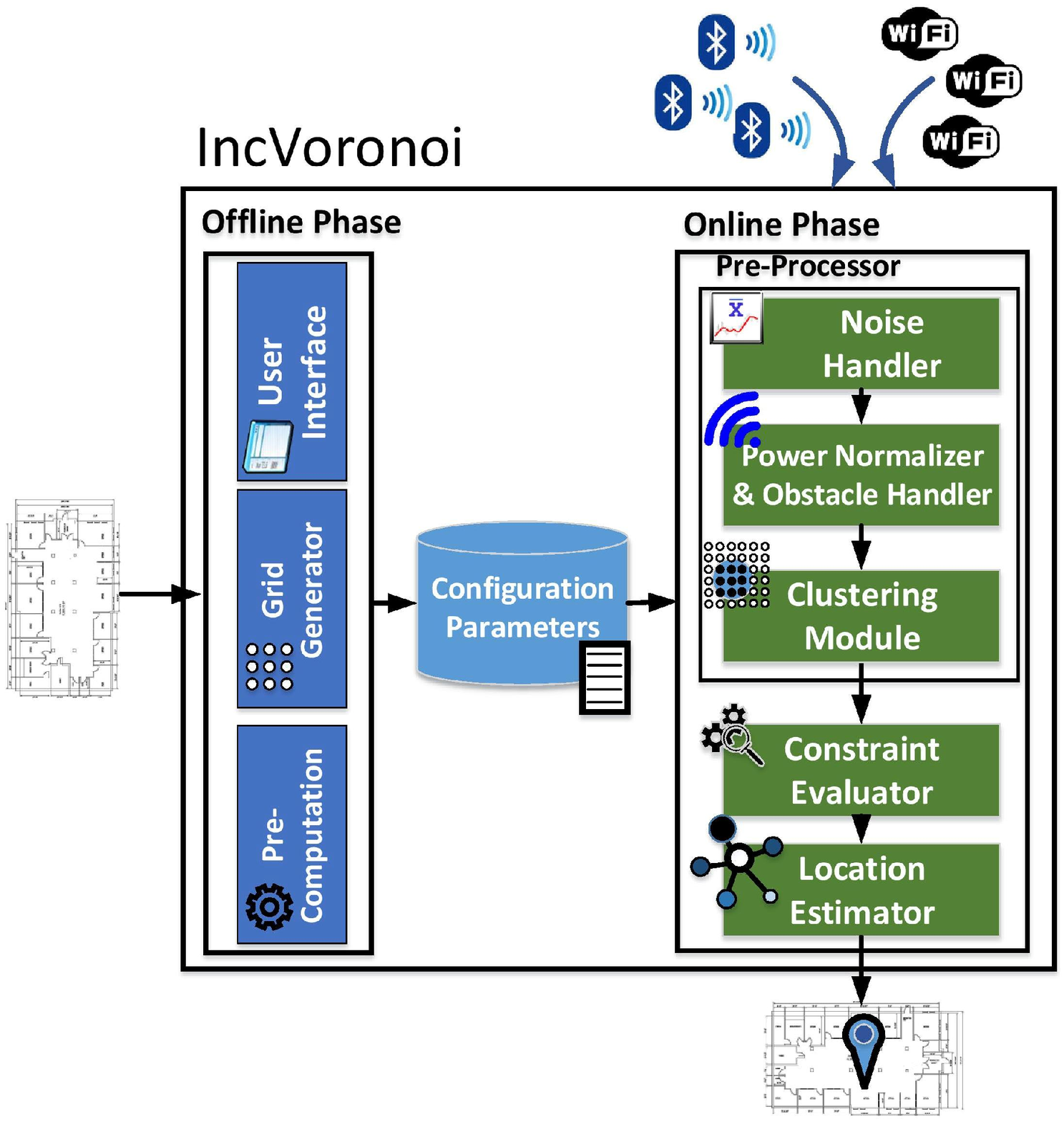}
\caption{\sys{} architecture.}
\label{fig:arch}
\end{figure}

\subsection{Architecture Overview}
Figure~\ref{fig:arch} shows the system architecture. The system works in two phases: an offline phase and an online tracking phase.

\subsubsection{Offline phase} During the offline phase, the system administrator uses the \textbf{\emph{User Interface}} module to enter a floorplan of the area of interest tagged with the APs locations and the walls. This last information can be obtained from the building CAD information or automatically constructed using crowdsourcing approaches as in \cite{crowdinside,elhamshary2014checkinside,elhamshary2015semsense,elhamshary2016transitlabel}; and is used by the \textbf{\emph{Power Normalizer and Obstacle Handler}} module during the online phase to reduce the impact of the building structure on the system accuracy. The virtual grid is also generated in this phase using the \textbf{\emph{Grid Generator}} module. The density of the grid can be configured by the system designer to trade off accuracy and computation overhead. The \textbf{\emph{Pre-Computation}} module finally calculates the associated parameters with each grid point (such as expected pairwise constraints evaluation based on the distance between the grid point and each pair of APs). This helps in reducing the running time during the online tracking phase. The \emph{Pre-Computation} module also calculates the Voronoi diagram of the area of interest. This is used for calculating the initial user ambiguity area as well as clustering the grid points during the online phase to reduce the computation overhead.

Note that \textbf{\emph{no calibration/war driving}} of the area of interest is performed in this phase. 
\subsubsection{Online tracking phase}
This is the main operational phase of \sys{}, where users can be tracked in realtime. The process starts by scanning for the APs that can be heard at the current device location. The \textbf{\emph{Pre-Processor}} compensates for the dynamic power control of the different APs and the obstacles in the environment as well as smoothes the input signal to reduce the wireless channel noise. It also selects a subset of the grid points to evaluate, further reducing the computational overhead. The \textbf{\emph{Constraints Evaluator}} then compares the expected RSS constraints with the stored distance constraints for each pair of heard APs for each grid point. The output of this module is a list of grid points and their associated number of matched constraints. Finally, the \emph{\textbf{User Location Estimator}} fuses the output of the \textbf{\emph{Constraint Evaluator}} module to obtain a single user location. 

For the rest of this section, we give the details of the main modules of the system.

\subsection{Pre-Processor}
The pre-processor has four goals: a) compensating for the dynamic power control of the different APs, b) taking the effect of the obstacles in the environment into account, c)~handling the noise in the wireless channel, and d) selecting a subset of the grid points to evaluate in the rest of the modules.

\subsubsection{Handling different transmit powers by the APs}
 To address this issue, \sys{} leverages the information available from the different RF technologies. Specifically, smart APs usually give access to their configuration information that can provide the current transmit power of the different APs. Similarly, the iBeacons technology includes the transmit power in the transmitted frames and this information is available from the APIs of mobile operating systems. \sys{} leverages this information to remove the transmit power offset by subtracting it from the scanned RSS vector.

\subsubsection{Handling obstacles in the environment}\label{sec:waf}
\begin{figure}[!t]
\centering
\includegraphics[width=0.77\columnwidth]{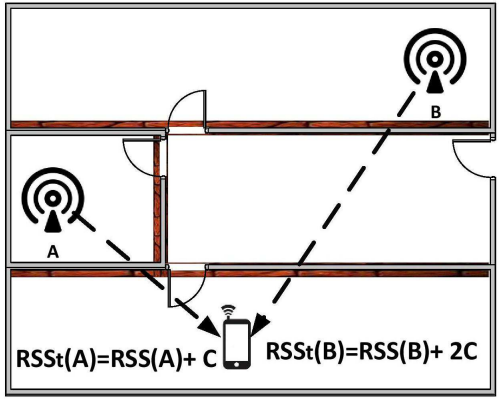}
\caption{Handling wall effect on the received signal strength.
}
\label{fig:WAF_comp}
\end{figure}

For a certain grid point and two specific APs, different number of walls can exist between this grid point and the two APs (Figure~\ref{fig:WAF_comp}). Since each wall leads to attenuation of the signal that goes through it, this leads to violating the homogeneous deployment assumption between the different APs. To address this issue, \sys{} estimates the number of walls (\emph{Walls}$(i, j)$) between each grid point $i$ and AP $j$ in its vicinity using the \emph{Pre-Computation} module during the offline phase. During the online tracking phase, the \emph{Constraints Evaluator} module uses the well-known wall attenuation factor model \cite{RADAR00} to compensate for the number of walls between an AP and a grid point. In particular, it transforms the input RSS scanning vector $s$ by adding a constant factor for each wall between the AP and the grid point. More formally, for grid point $i$ and AP $j$, the transformed RSS value (RSS$(j)_t$) is obtained from the input RSS (RSS$(j)$) as:
\begin{equation}
  \textrm{RSS}(j)_t=  \textrm{RSS}(j)+ \textrm{Walls}(i, j)\times \textrm{C}
\end{equation}
where $C$ is a constant parameter representing the signal attenuation due to a single wall. We evaluate the effect of $C$ on performance in Section~\ref{sec:eval}.

This transformation is applied to all APs heard at a specific grid point, unifying the effect of the different number of walls between them (Figure~\ref{fig:WAF_comp}).

\subsubsection{Noise reduction}
\label{sec:noise}

To help reduce the noisy effect of the wireless channel that may lead to swapping the constraints of two APs with close RSS, the pre-processor analyzes a window $w$ of RSS samples, rather than a single RSS value. Different filters can be used including the average and median filters. We also present a novel probabilistic approach for signal smoothing as part of the \emph{Constraints Evaluator} module. 
We show the effect of the different filters on performance in Section~\ref{sec:eval}.
\subsubsection{Grid points clustering}
To reduce the computational requirement of \sys{}, the pre-processor also selects a subset of the grid points for constraints evaluation. The idea is that the final user ambiguity region (and hence the most probable grid points) will be a subset of the Voronoi cell with the strongest heard AP. The pre-processor selects the points inside this Voronoi cell as the initial grid points for evaluating the RSS pairwise constraints. 

\subsection{Constraints Evaluator}
This module compares the expected pairwise distance constraints calculated during the offline phase with the actual RSS constraints in realtime for each selected grid point from the \emph{Pre-Processing} module. It outputs, for each grid point, the number of matched expected-actual constraints. 

Due to the noisy non-deterministic wireless channel, it is better to use a probabilistic approach for comparing the RSS of each pair of APs. Instead of relying on the average or median RSS only as in the \emph{Pre-Processing} module, \sys{} calculates the probability that the RSS received from one of the APs within a window of samples is higher than the other. This is based on the RSS histograms within a time window $w$. More formally, for two APs $A$ and $B$, $Pr(\textrm{RSS}_A>\textrm{RSS}_B)$ can be calculated as:
\begin{equation}
\label{eq:probMethodx}
Pr(\textrm{RSS}_A>\textrm{RSS}_B)=\sum\limits_{i=-\infty}^\infty (\sum\limits_{j=-\infty}^i Pr(\textrm{RSS}_B=j))\times Pr(\textrm{RSS}_A=i)
\end{equation}
If $Pr(\textrm{RSS}_A>\textrm{RSS}_B)$ is greater than 0.5, we consider AP $A$ to be the strongest AP. Otherwise, AP $B$ is considered to have a higher RSS. We compare this method to the different smoothing methods in the evaluation section.
\subsection{User Location Estimator}\label{sec:locest}
The \emph{User Location Estimator} aims to estimate the final user location from the candidate weighted grid points (weighted by the number of matched pairwise constraints). To do this, \sys{} uses the center of mass of the grid points that have the maximum number of matching constraints. In a perfect environment, these should correspond to the grid points within the final user ambiguity area.

To further enhance accuracy, the \emph{User Location Estimator} averages the last $t$ estimates to obtain a smoothed output.

\section{Performance Evaluation}
\label{sec:eval}

In this section, we evaluate the performance of \sys{} in a university campus building. We start by describing the experimental testbed and data collection process followed by evaluating the effect of the different parameters on performance and a comparison with two widely used fingerprinting-based and calibration-free techniques.

\begin{figure}[!t]
 \centering
    \includegraphics[width=0.9\columnwidth] {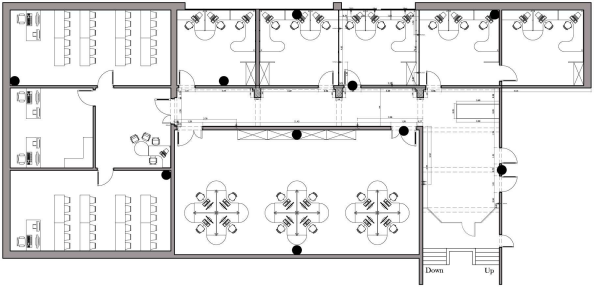}
    \caption{Testbed: Campus bldg (1 beacon/44m$^2$).}
    \label{fig:testbed1}
  \end{figure}

\subsection{Experimental Testbed and Data Collection}
 We deployed \sys{} in the second floor of our university campus (Figure~\ref{fig:testbed1}). Our experiments involved the iBeacons technology to avoid disrupting the operational WiFi network within our buildings when changing the parameters. The testbed spans an area of 26m$\times$17m covering corridors, offices, labs, and classrooms. The ceiling height is about 3m. Ten iBeacons were installed throughout the floor to provide uniform coverage as shown in Figure~\ref{fig:testbed1} with an average density of one beacon every 44m$^2$. 
All beacons are installed at the same height of 2.5 m. Table~\ref{tab:comp} summarizes the testbed parameters.

For evaluating the accuracy of the system, ground truth data was collected using Android phones. Test points were collected on a uniform grid with a 1m spacing. 
At each test point, beacons were scanned in each of the four directions: north, east, south, and west. 
 Note that \sys{} does not require any calibration. This data is for evaluating the system accuracy and for training the fingerprinting techniques. 

\begin{table}[!t]
\centering
\scalebox{1.2}{
\begin{tabular}{|l||l|l|} \hline
\textbf{Criteria}&\textbf{Value}\\ \hline \hline
Area ($m^2$)& 26$\times$17\\ \hline
Number of installed beacons&10\\ \hline
Beacon density ($m^2$/beacon)& 44\\ \hline
Building material& Brick\\ \hline
\hline
25\% accuracy (m)&1.5\\ \hline
Median accuracy (m)&2.8\\ \hline
75\% accuracy (m)&4.5\\ \hline
\end{tabular}}
\caption{Testbed parameters.}
\label{tab:comp}
\end{table}

\begin{table}[!t]
\centering
\scalebox{1.0}{
\begin{tabular}{|l||l|p{0.5in}|p{0.5in}|} \hline
\textbf{Parameter}&\textbf{Range}&\textbf{Default value}\\ \hline \hline
Preprocessing window ($w$)& 0 - 9&7 (Probabilistic)\\ \hline
Postprocessing window ($t$)& 0 - 10&6\\ \hline
Grid spacing (cm) & 25 - 200&50\\ \hline
Wall attenuation factor ($C$) (dBm)& 0 - 8&5\\ \hline
\end{tabular}}
\caption{Default parameters values used in evaluation.}
\label{tab:par}
\end{table}

\subsection{Effect of Parameters on Accuracy}
We evaluate the effect of the different parameters on the system performance: noise handling technique, wall attenuation factor, beacon density, virtual grid density, 
 location estimation technique, and post-processing smoothing window. 
Table~\ref{tab:par} shows the default parameters values.

\begin{figure}[!t]
\centering
\includegraphics[width=0.77\columnwidth]{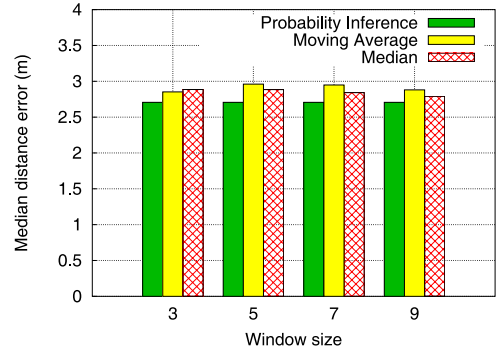}
\caption{Effect of the different noise handling techniques on accuracy.}
\label{fig:preprocess}
\end{figure}
\subsubsection{Noise handling technique}

Figure~\ref{fig:preprocess} shows the effect of the different noise handling techniques on the median distance error: the average, median, and the probabilistic constraint evaluator. The figure shows that the performance of all techniques enhances slightly with a larger window size. Both the median and average techniques have comparable performance. The proposed probabilistic constraint evaluator gives the best accuracy (from 5-8\% enhancement in median error) as it takes the full RSS histogram into account when evaluating the constraints compared to the median and average techniques that do not use the full available information.

\begin{figure*}[!t]
\noindent\begin{minipage}[t]{0.3\linewidth}
\centering
\includegraphics[width=1.1\columnwidth]{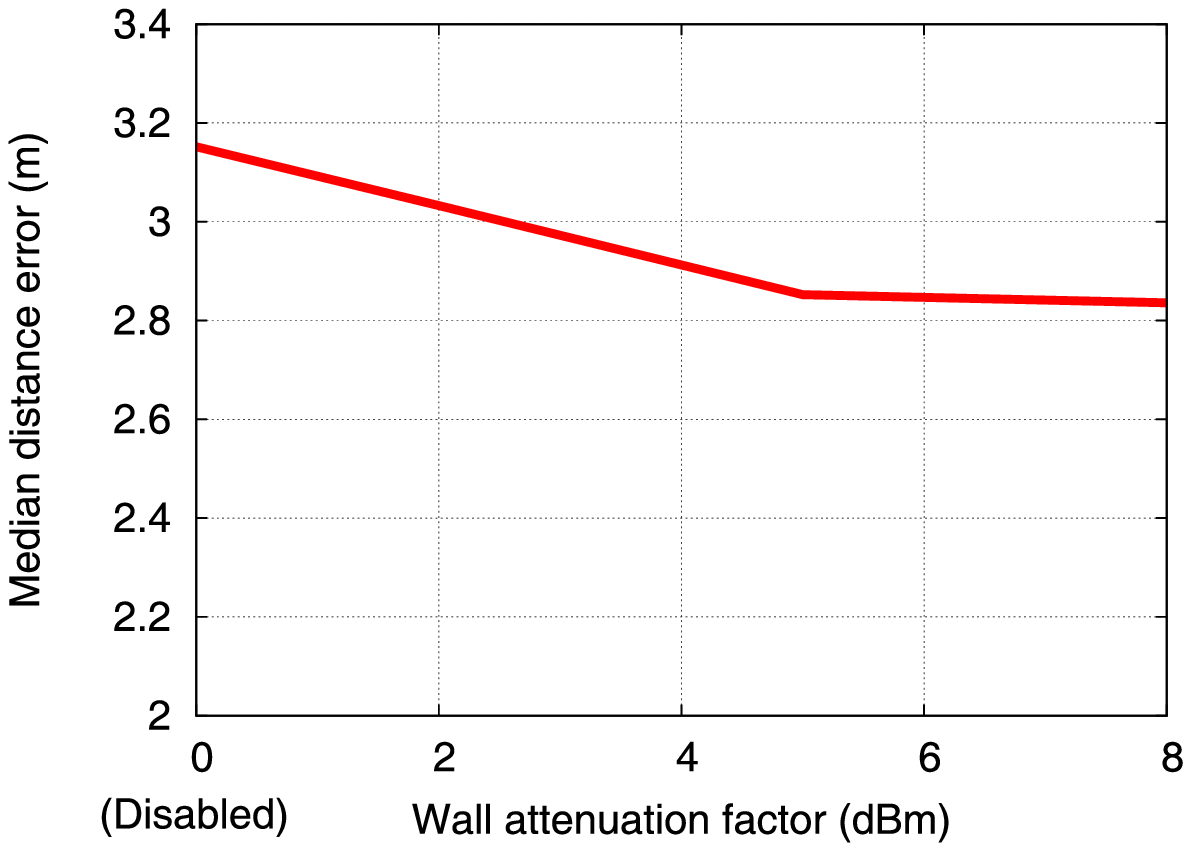}
\caption{Effect of the wall attenuation factor on accuracy.}
\label{fig:waf}
\end{minipage}
\hfill
\begin{minipage}[t]{0.3\linewidth}
\centering
\includegraphics[width=1.1\columnwidth]{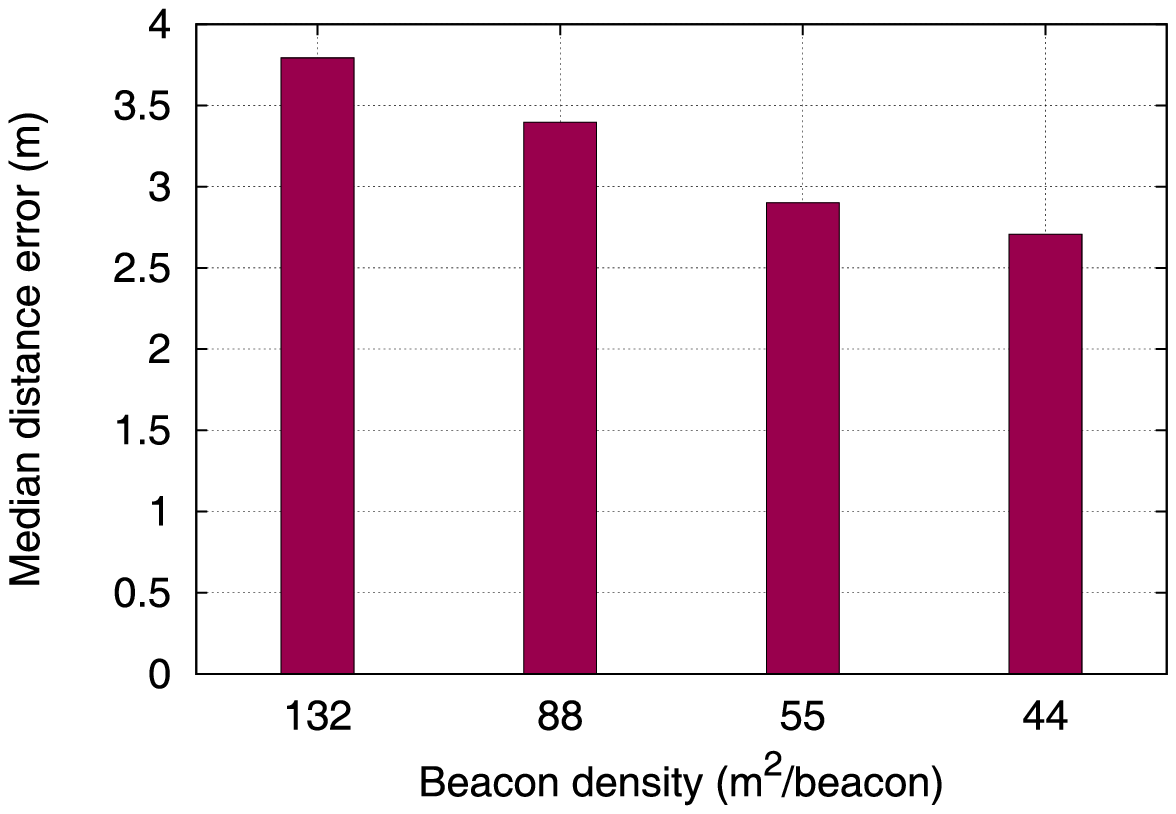}
\caption{Effect of the beacon density (area covered by each beacon/how far away to place each beacon) on accuracy.}
\label{fig:b_density}
\end{minipage}
\hfill
\begin{minipage}[t]{0.3\linewidth}
\centering
\includegraphics[width=1.1\columnwidth]{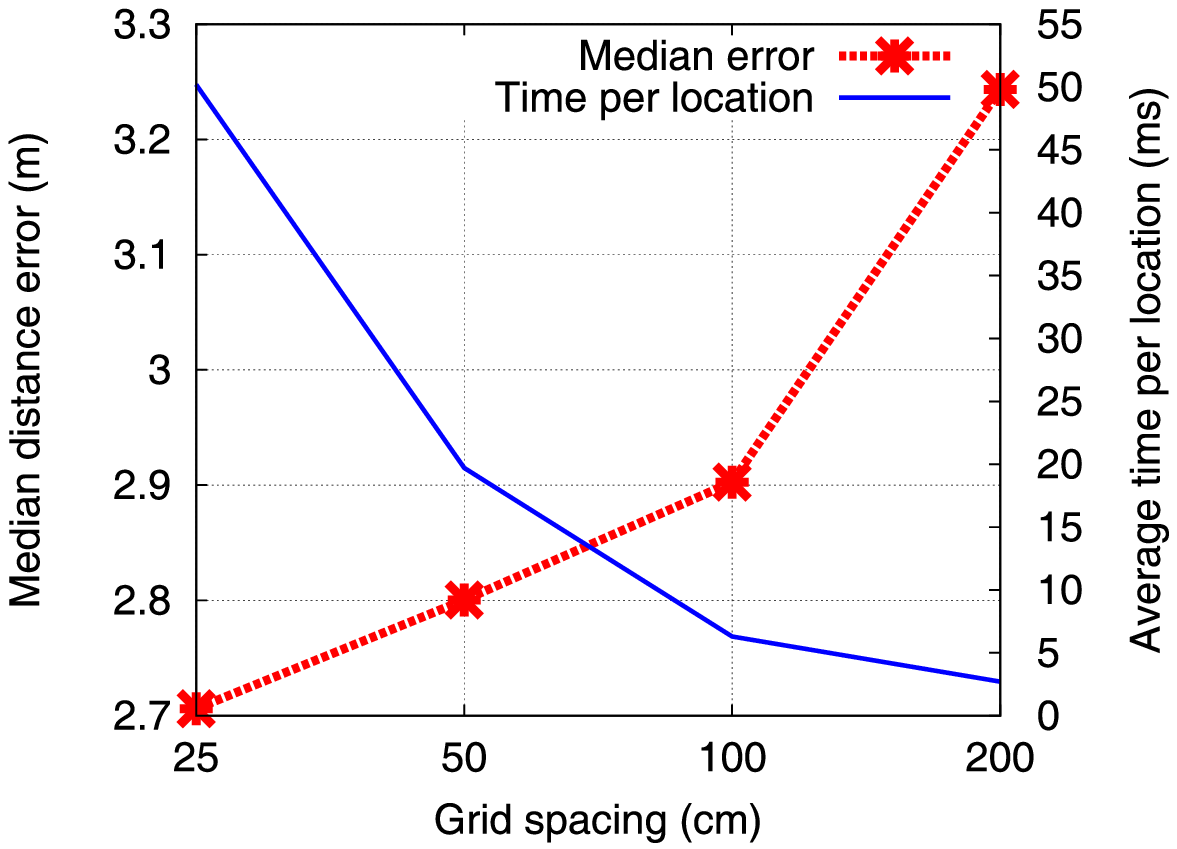}
\caption{Effect of the virtual grid density on system performance (Average time per location estimate on the \textbf{\emph{secondary axis}}).}
\label{fig:vgrid}
\end{minipage}
\end{figure*}

\subsubsection{Wall attenuation factor}
Figure~\ref{fig:waf} shows the effect of the wall attenuation factor used to capture the effect of the 
obstacles in the environment on the system accuracy. The figure shows that using the simple wall attenuation factor technique can enhance the accuracy by more than 10\% compared to the case of ignoring the obstacles in the environment.

\subsubsection{Beacon density}
Figure~\ref{fig:b_density} shows the effect of reducing the beacon density on accuracy. For this, we uniformly removed beacons from the total 10 installed in the area. The figure shows that even with a density as low as one beacon every 88 $m^2$, \sys{} can achieve a high accuracy of  less than 3.5m median distance error.
This highlights the promise of \sys{} for robust, high-accuracy, low-cost indoor localization. 

\subsubsection{Virtual grid density}
Figure~\ref{fig:vgrid} shows the effect of the virtual grid density on localization accuracy and average running time per location estimate.  The running time numbers are based on an HP Probook 450 laptop running a 2.5GHz Intel i5-4200M processor with 6 GB RAM. The figure shows that a grid size up to 1m is enough to maintain the high accuracy of \sys{}. Moreover, there is a natural tradeoff between accuracy and running time. However, even with a dense virtual grid of a 25cm spacing, the running time is only 50ms per location estimate. This is due to the clustering operation performed by the pre-processing module. 

\begin{figure}[!t]
\centering
\includegraphics[width=0.77\columnwidth]{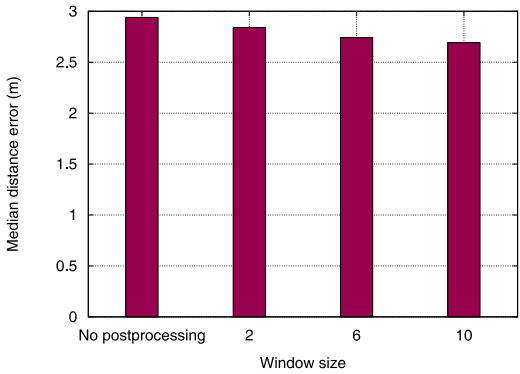}
\caption{Effect of the post-processing smoothing window on accuracy.}
\label{fig:postproc}
\end{figure}
\subsubsection{Post-processing smoothing window}
Figure~\ref{fig:postproc} shows that increasing the post processing window size slightly increases the performance. However, this comes with increased latency in response to user movement. A post-processing smoothing window of size $6$ balances the two effects.
\subsection{Comparison with Other Systems}
In this section, we compare the accuracy of \sys{} to probabilistic fingerprinting techniques (e.g. \emph{Horus} \cite{youssef2005horus}) 
 and another commonly used range-free technique under three changing operation conditions: Temporal changes, heterogeneous devices, and changes in the AP transmit power. In particular, we implemented and deployed 
a probabilistic fingerprinting technique \cite{youssef2005horus} and the weighted centroid technique \cite{blumenthal2007weighted} in the same testbed. For all of the three techniques, three test cases were evaluated against the \textbf{base fingerprint} (collected in July 
using a Samsung Galaxy S4 phone, and \mbox{-77dBm} transmit power): \textbf{a)} a test case collected in November 
with all other parameters fixed (temporal changes effect), \textbf{b)} a fingerprint with a Samsung Galaxy S4 Mini phone (heterogeneous devices effect), and \textbf{c)} a fingerprint with beacons transmit power set to -59 dBm (changes in AP transmit power effect). 
The fingerprint points were spaced 2m apart. Note that \sys{} does not require any fingerprinting overhead.

\begin{figure*}[!t]
\centering
    \subfloat[Temporal effect.\label{fig:fp_t}]{%
    \includegraphics[width=0.66\columnwidth]{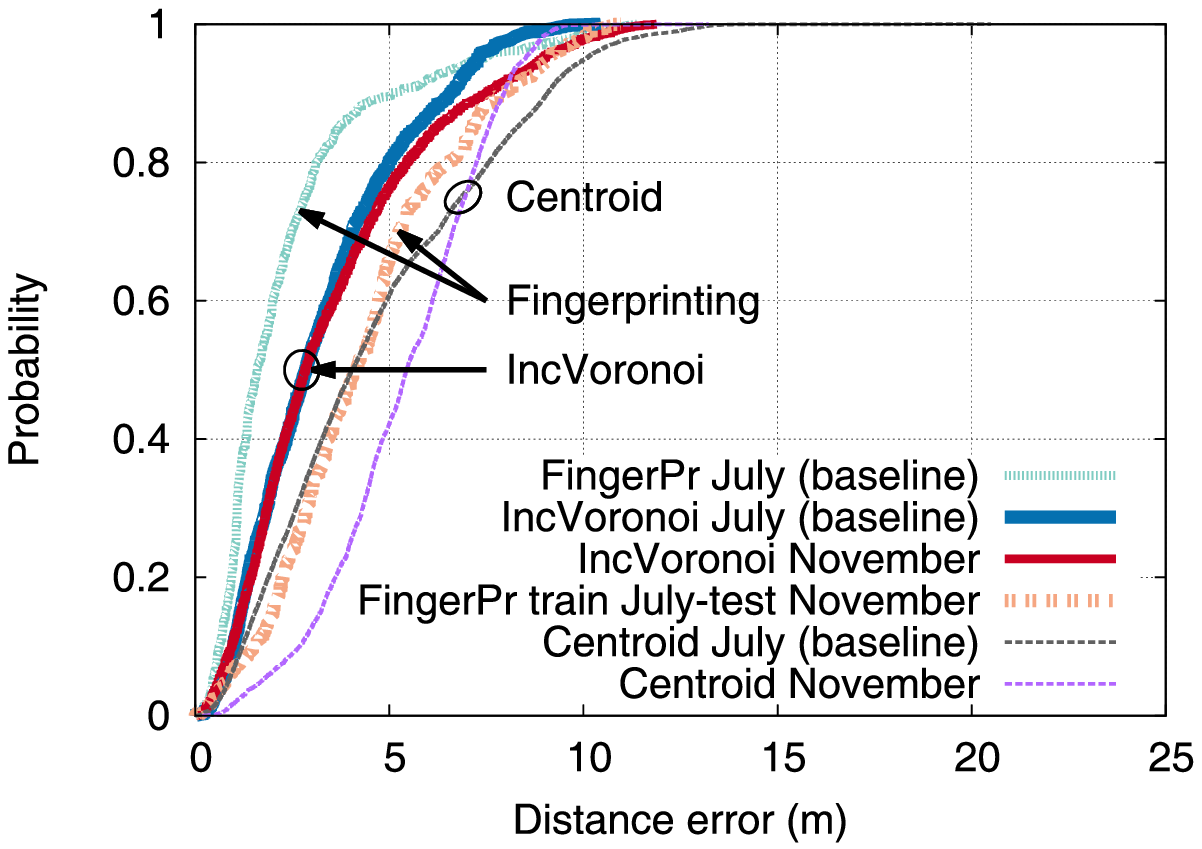}
    }
    \hfill
    \subfloat[Heterogeneous devices effect.\label{fig:fp_h}]{%
     \includegraphics[width=0.66\columnwidth]{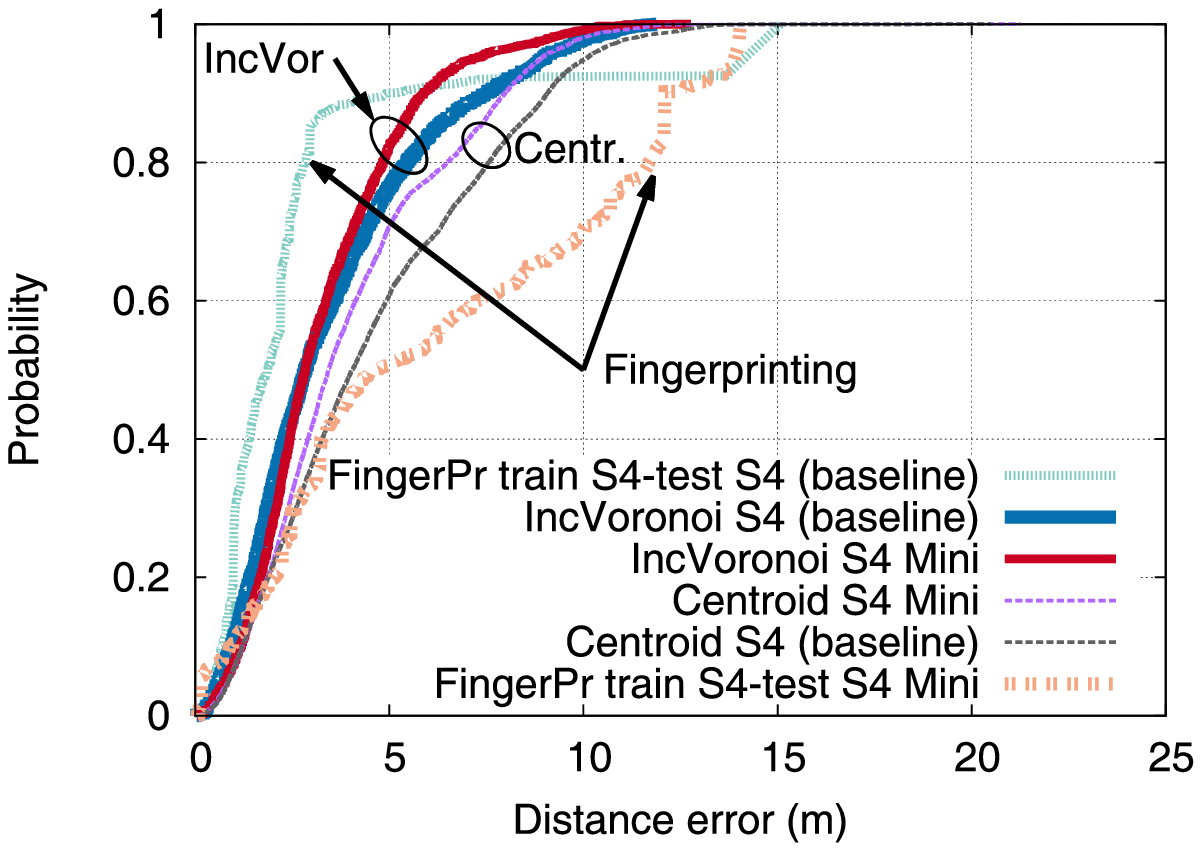}
     }\hfill
    \subfloat[Different AP transmit power effect.\label{fig:fp_a}]{%
     \includegraphics[width=0.66\columnwidth]{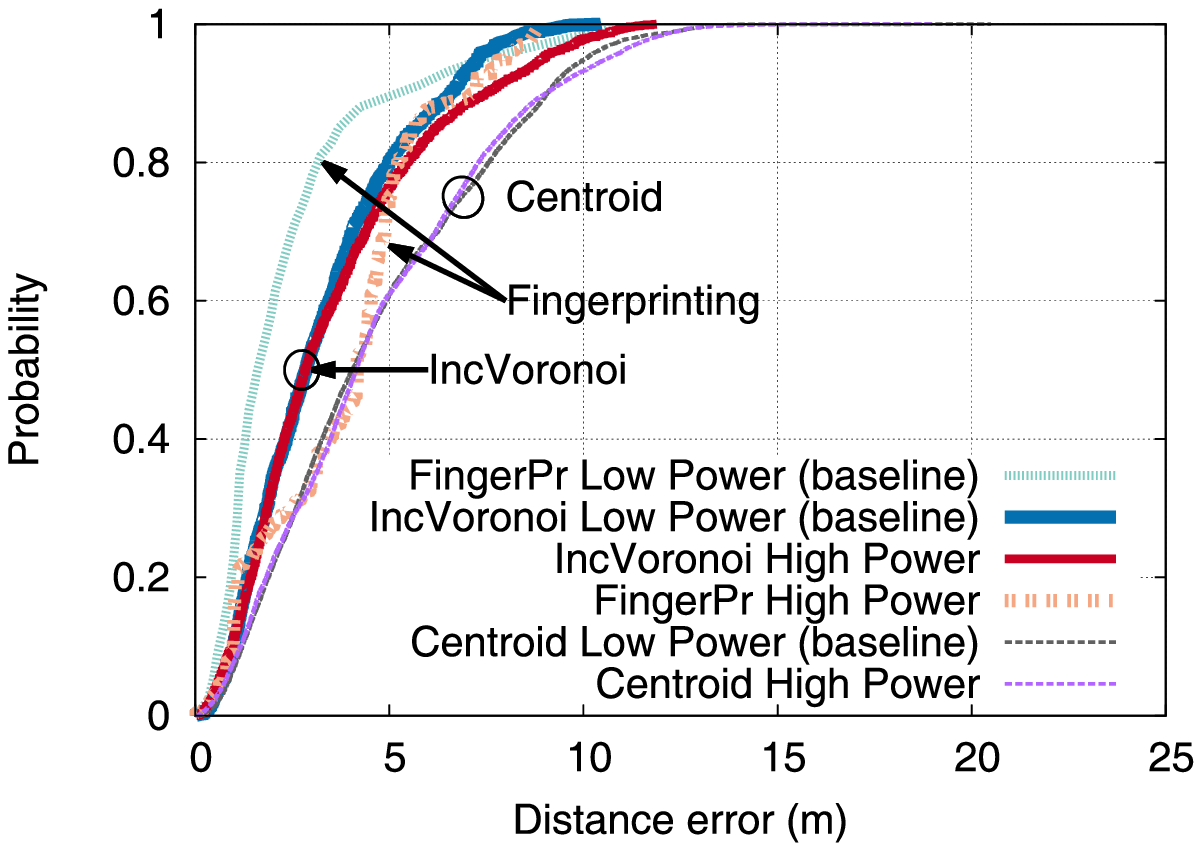}
     }
\caption{Performance comparison of \sys{} against the probabilistic fingerprinting and the centroid techniques under \textbf{three} operational environment changes. 
\sys{} performance is robust against changes compared to the other two systems.}
\label{fig:comp}
\end{figure*}
Figure~\ref{fig:comp} shows the CDF of distance error for the three techniques under different changes in the operation conditions. 
The figure shows that, even though the high overhead of the fingerprinting technique pays off as an advantage in accuracy when the fingerprint is ``\textbf{\emph{fresh}}'', this advantage quickly vanishes as the operation conditions change from those of the collected fingerprint.
The centroid technique can deal with changes in the transmit power well, similar to \sys{}. However, its performance degrades in the temporal and heterogeneous devices cases by at least 35\% compared to \sys{} as it cannot adapt to changes in the environment.

\emph{In summary - as shown in Table~\ref{tab:cdfTimeComp} - \sys{}, performs better than both the fingerprinting and the centroid techniques; it maintains its accuracy over time, with heterogeneous devices, and changes in the AP transmit power. Moreover, it does not require any calibration overhead.
}

\subsection{Discussion}\label{sec:locest}
The results in this section show that \sys{} can achieve a high accuracy of 2.8m median error \textbf{without any fingerprinting} with a low beacon density. This accuracy is robust to heterogeneity of users' devices, changes in AP transmit powers, and temporal changes in the environment.

The locations of the APs can be obtained from the buildings blueprints, manually tagged, or automatically estimated based on the RSS of the area of interest. Similarly, the floorplan of the area of interest is readily available for many buildings around the world or can be automatically inferred using recent systems, e.g. \cite{crowdinside,elhamshary2014checkinside,elhamshary2015semsense,elhamshary2016transitlabel}; all leading to \emph{virtually zero setup cost for the system}

The wall attenuation factor constants can be obtained automatically, based on the building material, from the CAD drawings or a pre-built material database. This highlights \sys{}'s\emph{ robustness to changes of testbeds and insensitivity to parameters changes}.

\section{Related Work}
\label{sec:related}

In this section, we focus on three groups of related work: systems that target reducing the RF fingerprint construction overhead, range-based 
 systems that eliminate the fingerprint creation yet have other disadvantages, and systems that address devices heterogeneity.

\begin{table*}[!t]
\centering
\begin{tabular}{|l|p{1.55in}||p{1.2in}|p{1.5in}|p{0.9in}|} \hline
\textbf{System}&\centering \textbf{\normalsize{\underline{Baseline}} \mbox{\small (Jul, Sam. S4, -77 dBm)}}&\centering  \textbf{Temporal change (Nov)}&\centering \textbf{Heterogeneous devices (S4Mini)}& \textbf{Diff. AP Tx power \mbox{\hspace{7mm}(-59dBm)}}\\ \hline \hline
\textbf{IncVoronoi}&\multicolumn{4}{c|}{2.8m}\\ \hline
\textbf{Fingerprinting} \cite{youssef2005horus},\cite{7103024}&\centering 1.6m&\centering  4.1m (\textbf{\textbf{-156}}\%)&\centering  4.6m (\textbf{\textbf{-187}}\%)&\multicolumn{1}{c|}{4.1m (\textbf{-156}\%)}\\ \hline
\textbf{Centroid} \cite{blumenthal2007weighted}&\centering 4.0m&\centering 5.4m (\textbf{-3}5\%)&\centering 3.4m (\textbf{-40}\%)&\multicolumn{1}{c|}{4.1m (\textbf{-2.5}\%)}\\ \hline
\end{tabular}
\caption{Comparison of the median error for different cases of operational changes. Numbers between parenthesis represent the degradation relative to the baseline of each technique.}
\label{tab:cdfTimeComp}
\end{table*}

\subsection{Reducing Fingerprint Construction Overhead}
A number of RF localization techniques have been proposed that try to reduce the calibration overhead through crowd-sourcing of the fingerprint construction \cite{abdelnassersemanticslam,elhamshary2016transitlabel,aly2014map++,elhamshary2015semsense,aly2015lanequest,aly2013dejavu
}, some through explicit \cite{organic} and others through implicit \cite{abdelnassersemanticslam,wang2012no,rai2012zee,aly2015lanequest,aly2013dejavu
} user feedback, using CAD tools or propagation models to automate the fingerprint construction process \cite{RADAR00,ARIADNE,aroma}, or combine RF localization with other sensors \cite{abdelnassersemanticslam,wang2012no,jiang2012ariel}. In \cite{organic}, the system detects areas with low fingerprint coverage and prompts the user to collect the fingerprint for them.
 SemanticSLAM \cite{wang2012no,abdelnassersemanticslam} and Zee \cite{rai2012zee} both leverage inertial sensors to get a rough estimate of the user location through dead-reckoning and associate a fingerprint with it. To reduce the error accumulation with dead-reckoning, Zee performs map matching with the floorplan while SemanticSLAM leverages unique ``anchors'' in the environment, detected based on the sensors signature, as error resetting opportunities. 
All these techniques require user participation. Therefore, their success depends on the user incentive system in place. In addition, explicit user participation may annoy the user while the accuracy of inertial sensors-based techniques depends on the  phone position and orientation \cite{mohssen2014s}, which is still an active area of research.

Other systems, e.g. \cite{RADAR00,ARIADNE,aroma}, use propagation models to estimate the fingerprint in the area of interest 
 rather than measuring it. The Radar system~\cite{RADAR00} uses a standard free-space propagation model and augments it with wall attenuation calculations to handle the complex indoor propagation conditions.
The ARIDANE~\cite{ARIADNE} and Aroma~\cite{aroma} systems use ray tracing models to get better RSS estimation in 2D and 3D respectively. These systems, however, still require samples from the environment to calibrate the model, require high computational requirements for ray tracing, 
 and the model parameters still depend on the specific phone used for measurements.

\emph{\sys{}, in contrast, requires neither user participation nor calibration measurements, and handles heterogeneous devices naturally.} 
\subsection{Range-based Systems} 
Range-based systems, typically used in sensor networks, e.g. \cite{youssef2007taxonomy,sheu2008distributed,ouadjaouteffective,yedavalli2005ecolocation,chen2009improved}, attempt to completely eliminate the dependence on a fingerprint by solving the localization problem geometrically. 
 To accomplish this, however, it is assumed in~\cite{sheu2008distributed} that the radio propagation model is a perfect circle which is unrealistic and leads to errors. To address this issue, the system proposed in~\cite{ouadjaouteffective} uses a a half-symmetric lens primitive while \cite{yedavalli2005ecolocation,chen2009improved} depend on the sensors sequence pattern. 
 However, these systems can only be applied to homogeneous networks like wireless sensor networks, where it is assumed that all sensors have the same properties as well as that all nodes can hear and communicate with each other, which is not the case with the passive iBeacons. In addition, they are evaluated through simulations. \emph{\sys{}, on the other hand, handles different transmit powers 
  and can work with nodes that do not interchange signals, which is the typical case of the iBeacons and WiFi APs. Moreover, it is designed to be robust to the dynamics of realistic environments.
}
\subsection{Heterogeneity Handling Techniques}
 To handle the devices heterogeneity, a number of approaches have been proposed that either map the fingerprint constructed by one device to another~\cite{het1,het2,het3} or use features that are device-independent~\cite{het5,het4}. For the first category, linear~\cite{het1}, non-linear~\cite{het2}, and probabilistic~\cite{het3} mappings have been applied with different accuracies. Device-independent features approaches either use specific features, e.g. the ratio between different APs RSS~\cite{het5}, or try to learn the device independent features automatically (e.g. by applying latent multi-task learning to labeled data ~\cite{het4}).

These techniques, however, require calibration between the different devices and the mapping function may not always be accurate, affecting the technique's accuracy. \emph{\sys{} on the other hand, depends on the relative relation of RSS with distance, which is device-independent and therefore does not require any calibration.
}

\section{Conclusion}
\label{sec:conclude}
We presented the design, implementation, and evaluation of \sys{}, a robust  accurate fingerprint-less indoor localization system that works in realistic deployment environments. \sys{} leverages a novel incremental Voronoi tessellation approach that reduces the user ambiguity region by applying successive RSS constraints.  We described the basic approach as well as different modules to handle practical deployment issues such as the noisy wireless channel, obstacles, dynamic power control, heterogeneous devices 
as well as efficient computations.

  Evaluation of \sys{} in a university campus building using the iBeacons technology on different Android phones shows that it can achieve a median distance error of 2.8m with a low beacon density of one beacon per 44$m^2$ area. Compared to the traditional fingerprinting technique, whose accuracy degrades by at least 156\%,  \sys{} is robust to temporal changes in the environment as well as changes in the devices and AP transmit power. 

  Currently, we are expanding \sys{} in different directions including deployment with different technologies, fusing with other sensors, combining with indoor map matching, among others.

\section{Acknowledgment}
This work is supported in part by NaviRize Inc. and in part by a grant from the Egyptian Information Technology Industry Development Agency (ITIDA).
\bibliographystyle{IEEEtran}

\end{document}